\documentclass[useAMS,usenatbib,twocolumn]{mn2e}
\usepackage[dvips]{graphicx}
\usepackage{amsmath}
\usepackage{amsfonts}
\usepackage{amssymb}

\title[How to Weigh a Star Using a Moon]{How to Weigh a Star Using a Moon}
\author[David M. Kipping]{David M. Kipping$^{1,2}$\thanks{E-mail:
d.kipping@ucl.ac.uk}\\
$^{1}$Department of Physics and Astronomy, University College London, Gower St., London WC1E 6BT \\
$^{2}$Harvard-Smithsonian Center for Astrophysics, 60, Garden St., Cambridge, MA 02138, USA}
\begin{document}

\date{Accepted 2010 September 21. Received 2010 September 17; in original form 2010 August 24}

\pagerange{\pageref{firstpage}--\pageref{lastpage}} \pubyear{2010}

\maketitle

\label{firstpage}

\begin{abstract}

We show that for a transiting exoplanet accompanied by a moon which also transits, the absolute masses and radii of the star, planet and moon are determinable. For a planet-star system, it is well known that the density of the star is calculable from the lightcurve by manipulation of Kepler's Third Law. In an analogous way, the planetary density is calculable for a planet-moon system which transits a star, and thus the ratio-of-densities is known. By combining this ratio with the observed ratio-of-radii and the radial velocity measurements of the system, we show that the absolute dimensions of the star and planet are determinable. This means such systems could be used as calibrators of stellar evolution. The detection of dynamical effects, such as transit timing variations, allows the absolute mass of the moon to be determined as well, which may be combined with the radius to infer the satellite's composition. 

\end{abstract}


\section{Introduction}\label{sec:intro}

Determining the absolute dimensions of a transiting planet system remains one of the most challenging problems confronting the exoplanetary scientist. The most widely adopted approach is to exploit Kepler's Third Law and the lightcurve morphology to determine the stellar density, $\rho_*$ \citep{sea03}. Combining $\rho_*$ with spectroscopic inferences of the surface gravity and metallicity allows us to produce stellar evolutionary tracks which may be used to estimate the mass and radius of the star (e.g. \citet{yi:2001}, \citet{baraffe:1998}). This approach is somewhat unsavoury as it is highly model dependent, where the model is that of stellar evolution.

If the spectral lines from the transiting body are resolvable, then the absolute dimensions may be determined as regularly done in the eclipsing binary community for double-lined, detached, eclipsing binary systems (DDEBs) (\cite{andersen:1991}; \cite{torres:2009} and references there-in). This approach was also recently used for the planet HD 209458b \citep{snellen:2010} but resulted in an error of greater than 20\% on the stellar mass. Another method is to use interferometry to measure the stellar radius but this is limited to nearby, bright targets \citep{baines:2007}.

A more subtle opportunity for determining the absolute dimensions was offered by \citet{agol:2005}, who considered dynamical interaction between two planets, where one transited the host star. The presence of the other planet induces transit timing variations (TTV) onto the eclipsing body (see also \citet{holman:2005}). By measuring both the TTV and lightcurve derived stellar density, a unique solution for the stellar mass is achievable. For a planet-moon system, perturbation of the planetary motion also gives rise to TTV \citep{sar99}, as well as transit duration variations (TDV) \citep{kip09a}.

In this letter, we show that the absolute masses and radii of the star and planet and the absolute radius of the moon are determinable for a transiting planet-moon system to an excellent approximation, even without measurable perturbation to the planetary orbit. The detection of dynamical effects from the moon, such as TTV, allows for the absolute mass of the moon to be found too and a more accurate solution for the dimensions of all three bodies. It is expected that the \emph{Kepler Mission} \citep{borucki:2009} will be sensitive to moons of $\gtrsim 0.2 M_{\oplus}$ \citep{kip09c} and thus such systems could be detected in the near future. These systems therefore would obviate the need for stellar evolutionary models and potentially act as evolutionary calibrators.

\section{Observables}
\subsection{Star-Planet System}

We start by considering the simple case of a planet-star system. The lightcurve is completely described by just a few free parameters. These are the time of mid-transit, $t_{\mathrm{mid}}$, the impact parameter, $b$, the orbital period, $P$, the ratio-of-radii, $p = R_P/R_*$ and the semi-major axis in units of the stellar radius, $a/R_*$. For eccentric systems we require the eccentricity, $e$, and the argument of periapsis, $\omega$, as well. In total, we have 7 free parameters. We note that time of mid-transit is essentially just a phase term.

Although $b$ and $a/R_*$ share a strong correlation \citep{car08}, typically $a/R_*$ is determinable to $\sim$10\% or better. For space based telescopes the precision can be a few percent, or better \citep{kipbak10b}. If we know $a/R_*$ and $P$, we may exploit Kepler's Third Law to determine the stellar density, as first pointed out by \citet{sea03}.

\begin{equation}
\rho_* + p^3 \rho_P = \frac{3\pi (a/R_*)^3}{G P^2}
\end{equation}

A typical approach is to either assume $\rho_* \gg p^3 \rho_P$ or iterate through the stellar evolution models a few times to improve the estimation of the $p^3 \rho_P$ term.

\subsection{Star-Planet-Moon System}

The lightcurve of a planet-moon system has similar free parameters to that of the planet alone. We assume that in the frame of the barycentre of the planet-moon system, the orbits of both bodies are unaffected by the star. Therefore, both the planet and moon orbit the barycentre in Keplerian orbits, which in turn orbits the star with Keplerian motion. This can be considered as a nested two-body simplification of the more general three-body problem. 

In this work, the only dynamical statement we make is that Kepler's Third Law is satisfied in these nested frames over the time-scale of the orbital period of the bodies. The validity of this approximation is discussed in the appendix, where we find that systems satisfying $f^3\ll(3/2)$ are well-described by our approximation (where $f$ is the semi-major axis of the moon around the planet in units of Hill radii). In practice, this means that all moons with $f\lesssim 0.5$ will be well-described here, which is known to be the bounding limit of prograde satellites \citep{dom06}.

In this set-up, the planetary parameters are unchanged, but should be given a subscript $_B$ to denote we are referring to the barycentre of the planet-moon system now. For example, $P$ becomes $P_B$, i.e. the orbital period of the planet-moon barycentre around the host star:

\begin{itemize}
\item[{\tiny$\blacksquare$}] $P_B$: Orbital period of the planet-moon barycentre around the star
\item[{\tiny$\blacksquare$}] $a_B/R_*$: Semi-major axis of the planet-moon barycentre around the star, in units of the stellar radius
\item[{\tiny$\blacksquare$}] $b_B$: Impact parameter of the planet-moon barycentre across the stellar face
\item[{\tiny$\blacksquare$}] $p$: Ratio of the planetary radius to the stellar radius
\item[{\tiny$\blacksquare$}] $t_{\mathrm{mid,B}}$: Time of mid-transit of the planet-moon barycentre across the stellar face
\item[{\tiny$\blacksquare$}] $e_B$: Eccentricity of the planet-moon barycentre around the star
\item[{\tiny$\blacksquare$}] $\omega_B$: Argument of periapsis of the planet-moon barycentre around the star
\end{itemize}

The satellite requires the same parameters except that it is understood we are referring to the orbit of the moon around the planet-moon barycentre at all times. The transit of a single body across the star is insensitive to the longitude of the ascending node but for two bodies the relative difference between the longitude of the ascending node becomes relevant. We therefore include one extra angle, $\Omega_S$ to account for this difference\footnote{The $_S$ subscript is used for the moon for consistency with previous works on exomoons e.g. \citet{simon:2007}}.

\begin{itemize}
\item[{\tiny$\blacksquare$}] $P_S$: Orbital period of the moon around the planet-moon barycentre
\item[{\tiny$\blacksquare$}] $a_S/R_*$: Semi-major axis of the moon around the planet-moon barycentre in units of the stellar radius
\item[{\tiny$\blacksquare$}] $b_S$: Impact parameter of the moon across the planetary face in the reference frame of the planet-moon barycentre.
\item[{\tiny$\blacksquare$}] $s$: Ratio of the moon's radius to the stellar radius
\item[{\tiny$\blacksquare$}] $\phi_{S}$: Phase angle between the time of the planet-moon barycentre's transit across the stellar face and the moon's transit across the star.
\item[{\tiny$\blacksquare$}] $e_S$: Eccentricity of the moon around the planet-moon barycentre
\item[{\tiny$\blacksquare$}] $\omega_S$: Argument of periapsis of the moon around the planet-moon barycentre
\item[{\tiny$\blacksquare$}] $\Omega_S$: Longitude of the ascending node of the moon around the planet-moon barycentre
\end{itemize}

As before, we expect $a_S/R_*$ and $b_S$ to share a strong correlation. Nevertheless, it should be expected that a unique solution to both $a_S/R_*$ and $P_S$ will be possible with sufficient quality photometry and quantity of transits. The investigation of parameter correlations and errors remains outside of the scope of this short letter.

It is easy to see how $a_S/R_*$ would reveal itself in the lightcurve though. For a moon on a circular orbit, $a_S/R_*$ would be revealed by the temporal separation between the moon's mid-transit time and the planet's mid-transit time. In practice, one would fully model the transit to include both bodies and fit for $a_S/R_*$.

The other critical parameter, $P_S$, is also known to be measurable. For example, \citet{kip09a} showed that the ratio of the TDV to TTV amplitude directly provides $P_S$. Another method is to use the time difference between exomoon transit features. For short period moons, this can be determined even with a single transit in some cases, as pointed out by \citet{sato:2009}.

\subsection{Dynamical effects}

The planetary motion is perturbed by the presence of the moon giving rise to transit timing variations (TTV) \citep{sar99} and transit duration variations (TDV) \citep{kip09a,kip09b}. To completely describe the planetary motion, we still require on extra parameter, the mass ratio $\mathcal{M}_{SP} = M_S/M_P$, which is directly determinable from timing effects. With this ratio, and all of the previously mentioned parameters, the lightcurve of a planet-moon system may be completely described.

\section{Densities}

Before we present our equations for determining the densities of all three bodies, we note that \citet{simon:2007} discuss how the ratio of the exomoon-to-exoplanet density may be determined using a modified definition of transit time variation (TTV) which the authors labelled TTV$_p$, or photometric TTV. However, the influence of the moon on the stellar and planetary derived densities was not considered and the need for a customized definition of TTV is obviated in the equations presented in this work. Also, the authors did not discuss determining the absolute dimensions of the star or planet.

\subsection{Planetary density}

As we saw earlier, the ability to be measure $a_B/R_*$ and $P_B$ for a transiting planet means that we can determine $\rho_*$ using Kepler's Third Law. In the same way, the ability to measure $a_S/R_*$ and $P_S$ for the moon-planet system allows us to determine $\rho_P$. Note that an explicit assumption throughout this work, as for \citet{sea03}, is that all bodies are perfect spheres.

One subtle difference is that the distance $a_B$ is defined from the planet-moon barycentre to the star, but $a_S$ is the moon to planet-moon barycentre separation. Kepler's Third Law is applicable to motion in the frame of reference of one of the two bodies, and not the barycentre. The planet-moon separation is therefore given by $a_{SP} = a_S (1+\mathcal{M}_{SP})$. Given that $a_S/R_*$ and $\mathcal{M}_{SP}$ are observables, then $a_{SP}/R_*$ is an observable too. Using Kepler's Third Law, we have:

\begin{align}
\frac{a_{SP}}{R_*} &= \frac{1}{R_*} \Big(\frac{G M_B P_S^2}{4\pi^2}\Big)^{1/3} \nonumber \\
\rho_P + \Big(\frac{s}{p}\Big)^3 \rho_S &= \frac{3 \pi (a_{SP}/R_*)^3}{G P_S^2 p^3}
\end{align}

For convenience, we replace $a_{SP}$ with $a_S$.

\begin{equation}
\rho_P + \Big(\frac{s}{p}\Big)^3 \rho_S = \frac{3 \pi (a_S/R_*)^3 (1+\mathcal{M}_{SP})^3}{G P_S^2 p^3} 
\end{equation}

The above expression is comparable to that used for deriving the stellar density in equation (1). However, (1) must be modified to account for the planet-moon system too:

\begin{align}
&\frac{a_B}{R_*} = \frac{1}{R_*} \Big(\frac{G (M_B+M_*) P_B^2}{4\pi^2}\Big)^{1/3} \nonumber \\
&\rho_* + p^3 \rho_P + s^3 \rho_S = \frac{3\pi (a_B/R_*)^3}{G P_B^2}
\end{align}

Unfortunately, both equations (3) and (4) involve other densities and are not written purely in terms of the observables. A resolution to this is to consider the satellite density and then return to these equations.

\subsection{Satellite density}

If we assume that one has measured $M_S/M_P$ from the lightcurve, due to dynamical effects, it possible to evaluate the density of the satellite from the lightcurve alone.

\begin{align}
\rho_S &= \frac{M_S}{(4/3)\pi R_S^3} \nonumber \\
\rho_S &= \Big(\frac{p}{s}\Big)^3 \rho_P \mathcal{M}_{SP}
\end{align}

Feeding this back in to the planetary density equation gives an equation for $\rho_P$ purely in terms of the observables.

\begin{equation}
\rho_P = \frac{3 \pi (a_S/R_*)^3 (1+\mathcal{M}_{SP})^2}{G P_S^2 p^3} 
\end{equation}

It should be stressed at this stage that it is possible to provide an excellent estimation of $\rho_P$ even if no detectable TTV/TDV effects exist by setting $\mathcal{M}_{SP} \ll 1$. Indeed, the fact that no dynamical effects are detected actually requires $\mathcal{M}_{SP} \ll 1$. However, we continue using the unapproximated forms and present $\rho_S$ in terms of the observables only:

\begin{equation}
\rho_S = \frac{3 \pi (a_S/R_*)^3 (1+\mathcal{M}_{SP})^2 \mathcal{M}_{SP}}{G P_S^2 s^3}
\end{equation}

And finally $\rho_*$:

\begin{equation}
\rho_* = \frac{3\pi}{G} \Bigg[ \frac{(a_B/R_*)^3}{P_B^2} - \frac{(a_S/R_*)^3 (1+\mathcal{M}_{SP})^3}{P_S^2}\Bigg]
\end{equation}

\section{Absolute Dimensions}

Armed with $p$, $\rho_*$ and $\rho_P$, it is possible to derive $\mathcal{M}_{P*} = M_P/M_*$.

\begin{align}
\mathcal{M}_{P*} = \frac{M_P}{M_*} &= p^3 \frac{\rho_P}{\rho_*}
\end{align}

In the case of $\mathcal{M}_{SP} \ll 1$, we have shown earlier how $\rho_P$ is derivable even without detecting any TTV/TDV effects. Therefore, $M_P/M_*$ is also derivable without TTV/TDV under the same approximation. However, continuing using the full unapproximated equation, it is possible to write the mass ratio of the planet-moon pair to the star as:

\begin{align}
\mathcal{M}_{B*} &= \mathcal{M}_{P*} + \mathcal{M}_{S*} \nonumber \\
\mathcal{M}_{B*} &= \Bigg[ \frac{P_S^2 (a_B/R_*)^3}{P_B^2 (a_S/R_*)^3} \frac{1}{(1+\mathcal{M}_{SP})^{3}} - 1 \Bigg]^{-1}
\end{align}

At this point, we invoke information from a non-lightcurve origin. The radial velocity semi-amplitude contains information on the masses in the system and may be used to break the degeneracy one would otherwise be stuck with. Taking the standard equation for the semi-amplitude of the RV signal, and accounting for the fact the reflex motion of the star is now due to the combined mass of the planet-moon pair, we have:

\begin{equation}
K_*^3 = \frac{2 \pi G M_B^3 \sin^3 i_B}{(M_* + M_B)^2 P_B (1-e_B^2)^{3/2}}
\end{equation}

Re-arranging for the stellar mass and replacing $M_B = M_* \mathcal{M}_{B*}$, we obtain:

\begin{align}
M_* &= \frac{(1 + \mathcal{M}_{B*})^2 P_B K_*^3 (1-e_B^2)^{3/2}}{2 \pi G \mathcal{M}_{B*}^3 \sin^3 i_B}
\end{align}

This may be re-expressed in terms of the observables only:

\begin{align}
M_* &= \Bigg[\frac{(a_B/R_*)^6}{(a_S/R_*)^9}\Bigg] \Bigg[\frac{P_S^4}{P_B^5}\Bigg] \Bigg[\frac{(1-e_B^2)^{3/2} K_*^3}{2 \pi G \sin^3 i_B}\Bigg] \nonumber \\
\qquad& \times \Bigg[ \frac{(a_B/R_*)^3 P_S^2-(a_S/R_*)^3 (1+\mathcal{M}_{SP})^3 P_B^2}{(1+\mathcal{M}_{SP})^9 } \Bigg]
\end{align}

Similarly, the stellar radius may be written as:

\begin{equation}
R_* = \frac{(a_B/R_*)^2 \sqrt{1-e_B^2} K_* P_S^2}{2 \pi \sin i_B (a_S/R_*)^3 (1+\mathcal{M}_{SP})^3 P_B}
\end{equation}

Once the stellar mass and radius is known, the mass and radius of the planet are easily found. Once again, all of these quantities are derivable to an excellent approximation even without TTV/TDV by simply using $\mathcal{M}_{SP} \ll 1$ in equations (13) and (14). However, one can see that in the absence of a moon we have $(a_S/R_*)/P_S\rightarrow 0/0$ and (13) \& (14) become meaningless, as expected. Given that $R_*$ and $s=R_S/R_*$ have been determined, then the absolute radius of the moon is also known. In the absence of any TTV/TDV, we would be able to place at least an upper limit satellite mass, which would constrain the internal composition of the moon. However, if dynamical effects are detected then the more accurate solutions in equation (13) and (14) may be used and the absolute dimensions of all three bodies are determinable (Table~1).

\begin{table}
\caption{\emph{Summary of derivables in two cases. Column 2 shows the case where no dynamical effects are detectable (e.g. TTV/TDV) but the planet and moon transit the host star. Column 3 shows the case where dynamical effects are detectable.}} 
\centering 
\begin{tabular}{c c c} 
\hline\hline 
Parameter & No TTV/TDV & TTV/TDV \\ [0.5ex] 
\hline 
$M_*$ & Approximate solution & Accurate solution \\
$R_*$ & Approximate solution & Accurate solution \\
$M_P$ & Approximate solution & Accurate solution \\
$R_P$ & Approximate solution & Accurate solution \\
$M_S$ & Upper limit only     & Accurate solution \\  
$R_S$ & Approximate solution & Accurate solution \\ [1ex]
\hline\hline 
\end{tabular}
\label{table:summary} 
\end{table}

\section{Conclusions}

We have shown that if the semi-major axis of a moon's orbit around an exoplanet and its corresponding orbital period are measurable, as is expected for a transiting planet-moon system, then it is possible to determine the densities of the star and planet. The detection of dynamical effects in the transit lightcurve (e.g. TTV/TDV effects) allows for an accurate determination of the various mass ratios and the density of the moon as well. However, an excellent approximate expression for the mass ratio of the planet-moon pair to the star is still derivable even in the cases where $M_S/M_P \ll 1$ and no detectable TTV/TDV effects occur.

Combining this mass ratio with the radial velocity data allows for an excellent approximate solution for the absolute masses and radii of the star and planet and the absolute radius of the satellite. The detection of dynamical effects allows for the more accurate solution for the absolute dimensions of all three bodies. The equations are not limited to strictly planet-moon cases but naturally are applicable to binary-planets as well, should they be detected \citep{podsiadlowski:2010}. 

Aside from such systems being interesting for harbouring a moon, they are also very valuable to validate the stellar evolution models, which are usually invoked for transiting planet systems. Planet-moon systems could therefore act as evolution calibrators, in the same way as double-lined, detached, eclipsing binaries (DDEBs) do. The solution also means that the absolute mass and radius of the planet and moon are resolvable. This is particularly useful for determining the composition of the moon. An icy moon (which may have become a water-world after planetary migration) would have a significantly lower density than a rocky planet which may have been captured \citep{grindrod:2007}. Therefore, the density determination alone allows for some assessment of the possible origin of the moon \citep{namouni:2010} and possibly even its habitability \citep{williams:1997}.

The ability to measure the absolute mass and radius of the moon is extremely powerful. The moon is likely to be sub-Earth mass and modern radial velocity measurements would not be able to detect a planet of equivalent mass \citep{lovis:2008}. Therefore, moons measured in this way could become the objects of smallest measurable mass outside of the Solar System.

\section*{Acknowledgments}

DMK has been supported by UCL, the Science Technology \& Facilities Council (STFC) studentships and the SAO predoctoral fellowships. Thanks to J. Irwin and the anonymous reviewer for useful comments.

\section*{Appendix}

We will here explore the conditions under which our nested two-body formulation is a good approximation to the three-body problem, which can only be solved exactly through numerical techniques. The motion of the satellite, as the lowest-mass object in our hypothetical system, will be the body which experiences the greatest departure from the simple nested two-body approximation. In this letter, the derivation has operated under the assumption that in the frame of reference of the planet, Kepler's Third Law applies for the motion of the moon, independent of the star. We are interested in the conditions under which the motion of the moon departs significantly from this statement (i.e. the star perturbs the satellite's orbit) and therefore our expressions are no longer valid\footnote{Orbital variations occurring on time-scales much greater than the orbital period, such as precession of the apsides or nodes, has a negligible impact on our derivation.}.

For a planet-moon pair at infinite orbital separation, the validity our assumption can be understood intuitionally. As the planet-moon pair moves in, the disturbance due to the star becomes greater and our approximation will break down. We may follow the standard treatment used in lunar theory \citep{mur99} to calculate the disturbance to our approximation. Let us select a non-rotating reference frame with the star at the rest at the origin, for which we may treat the frame as inertial since the mass of the star is much greater than the planet or moon. The satellite's equation of motion is given by:

\begin{align}
\ddot{\mathbf{r}_S} &= - n_S^2 a_S^3 \frac{(\mathbf{r}_S - \mathbf{r}_P)}{|\mathbf{r}_S - \mathbf{r}_P|^3} - n_P^2 a_P^3 \frac{\mathbf{r}_S}{|\mathbf{r}_S|^3}
\end{align}

Where $\mathbf{r}_S$ and $\mathbf{r}_P$ are the position vectors of the satellite and planet relative to the star, $n_P$ is the mean-motion of the planet around the star and $a_P$ is the corresponding semi-major axis. Similarly, $n_S$ is the mean-motion of the moon around the planet and $a_S$ is the corresponding semi-major axis. To translate to the reference frame of the planet, we may substitute:

\begin{align}
\mathbf{r}  &= \mathbf{r}_S - \mathbf{r}_P \\
\mathbf{r}' &= -\mathbf{r}_P
\end{align}

These vectors now describe the positions of the satellite and the star, respectively, relative to the planet. It follows that in this non-inertial frame where the planet is at rest, but the coordinate axes point in \emph{fixed} directions, the equation of motion for the satellite becomes:

\begin{align}
\ddot{\mathbf{r}} &= - n_S^2 a_S^3 \frac{\mathbf{r}}{|\mathbf{r}|^3} + n_P^2 a_P^3 \Bigg[ \frac{(\mathbf{r}'-\mathbf{r})}{|\mathbf{r}'-\mathbf{r}|^3} - \frac{\mathbf{r}'}{|\mathbf{r}'|^3}\Bigg]
\end{align}

The term to the right of the addition sign represents the disturbance to Kepler's Third Law from a simple nested two-body approximation. The disturbing function can be written as:

\begin{align}
\Phi &= n_P^2 a_P^3 \Bigg[ \frac{\mathbf{r}_S}{|\mathbf{r}_S|^3} - \frac{\mathbf{r}_P}{|\mathbf{r}_P|^3}\Bigg] \nonumber \\
|\Phi| &\simeq n_P^2 a_P \Big( [1 + (a_S/a_P)]^{-2} - 1\Big) \nonumber \\
|\Phi| &\simeq 2 n_P^2 a_S
\end{align}

Where have assumed a low eccentricity system and expanded to first order for $a_S/a_P \ll 1$. For the disturbing function to be small, we therefore require:

\begin{align}
\Big| n_S^2 a_S^3 \frac{\mathbf{r}}{|\mathbf{r}|^3} \Big| \gg |\Phi| \nonumber \\
n_S^2 \gg 2 n_P^2
\end{align}

Where the final line gives the condition under which our derivation is ultimately valid and is equivalent to:

\begin{equation}
P_S^2 \ll P_P^2/2
\end{equation}

It is possible to write $P_S$ in terms of $P_P$ and the moon's orbital distance in units of Hill radii, $f$, using the relation of \citet{kip09a} that $P_S/P_P \simeq \sqrt{f^3/3}$. Putting this together yields:

\begin{equation}
f^3 \ll 3/2
\end{equation}

Defining $\ll$ to indicate an order-of-magnitude difference, i.e. a factor of 10, this constrains $f\leq0.531$. We note that this distance is larger than that predicted as the maximum stable separation for a prograde satellite of $f=0.4895$ \citep{dom06}. However, a retrograde moon can be stable at up to $f=0.9309$. Under such conditions, the equations presented in this work would become invalid.

\end{document}